\documentclass[sigconf]{acmart}
\AtBeginDocument{%
  \providecommand\BibTeX{{%
    \normalfont B\kern-0.5em{\scshape i\kern-0.25em b}\kern-0.8em\TeX}}}

\setcopyright{acmcopyright}
\copyrightyear{2018}
\acmYear{2018}
\acmDOI{XXXXXXX.XXXXXXX}

\acmConference[Conference acronym 'XX]{Make sure to enter the correct
  conference title from your rights confirmation emai}{June 03--05,
  2018}{Woodstock, NY}
\acmBooktitle{Woodstock '18: ACM Symposium on Neural Gaze Detection,
 June 03--05, 2018, Woodstock, NY} 
\acmPrice{15.00}
\acmISBN{978-1-4503-XXXX-X/18/06}

\usepackage{booktabs}
\usepackage{multirow}
\usepackage{algorithm}
\usepackage{algorithmic}
\usepackage{enumitem}
\usepackage{xcolor}
\usepackage{subfigure}
\usepackage{diagbox} 

\newcommand{\eg}{\emph{e.g.}}
\newcommand{\ie}{\emph{i.e.}}

\newif\ifshowcomments
\showcommentstrue   

\ifshowcomments
    \newcommand{\jc}[1]{{\color{purple} [jiachen: #1]}}
    
    \newcommand{\ljh}[1]{{\color{red}[ljh: #1]}}
\else
    \newcommand{\jc}[1]{}
    \newcommand{\ljh}[1]{}
\fi

\begin{document}

\title{Lifelong Personalized Low-Rank Adaptation of Large Language Models for Recommendation}

\author{Jiachen Zhu}
\orcid{0000-0002-9140-1429}
\affiliation{%
    \institution{Shanghai Jiao Tong University}
    \city{Shanghai}
    \country{China}}
\email{gebro13@sjtu.edu.cn}

\author{Jianghao Lin}
\orcid{0000-0002-8953-3203}
\affiliation{%
    \institution{Shanghai Jiao Tong University}
    \city{Shanghai}
    \country{China}}
\email{chiangel@sjtu.edu.cn}

\author{Xinyi Dai}
\affiliation{%
    \institution{Huawei Noah's Ark Lab}
    \city{Shanghai}
    \country{China}}
\email{daixinyi5@huawei.com}

\author{Bo Chen}
\orcid{0000-0001-7053-8269}
\affiliation{%
    \institution{Huawei Noah's Ark Lab}
    \city{Shenzhen}
    \country{China}}
\email{chenbo116@huawei.com}

\author{Rong Shan}
\affiliation{%
    \institution{Shanghai Jiao Tong University}
    \city{Shanghai}
    \country{China}}
\email{shanrong@sjtu.edu.cn}

\author{Jieming Zhu}
\affiliation{%
    \institution{Huawei Noah's Ark Lab}
    \city{Shenzhen}
    \country{China}}
\email{jamie.zhu@huawei.com}

\author{Ruiming Tang}
\orcid{0000-0002-9224-2431}
\affiliation{%
    \institution{Huawei Noah's Ark Lab}
    \city{Shenzhen}
    \country{China}}
\email{tangruiming@huawei.com}

\author{Yong Yu}
\orcid{0000-0003-0281-8271}
\affiliation{%
    \institution{Shanghai Jiao Tong University}
    \city{Shanghai}
    \country{China}}
\email{yyu@apex.sjtu.edu.cn}

\author{Weinan Zhang}
\authornote{Weinan Zhang is the corresponding author.}
\orcid{0000-0002-0127-2425}
\affiliation{%
    \institution{Shanghai Jiao Tong University}
    \city{Shanghai}
    \country{China}}
\email{wnzhang@sjtu.edu.cn}

\renewcommand{\shortauthors}{Jiachen Zhu, et al.}

\begin{abstract}
We primarily focus on the field of large language models (LLMs) for recommendation, which has been actively explored recently and poses a significant challenge in effectively enhancing recommender systems with logical reasoning abilities and open-world knowledge. Current mainstream efforts mainly center around injecting personalized information from recommendation models into LLMs by customizing input templates or aligning representations between semantic and recommendation spaces at the prediction layer. However, they face three significant limitations: (1) LoRA is mostly used as a core component in existing works, but personalization is not well established in LoRA parameters as the LoRA matrix shared by every user may not cater to different users' characteristics, leading to suboptimal performance. (2) Although lifelong personalized behavior sequences are ideal for personalization, their use raises effectiveness and efficiency issues since LLMs require escalating training and inference time to extend text lengths. (3) Existing approaches fail to be scalable for large datasets due to training efficiency constraints. Thus, LLMs only see a small fraction of the datasets (e.g., less than 10\%) instead of the whole datasets, limiting their exposure to the full training space. To address these problems, we propose \textbf{RecLoRA}. This model incorporates a personalized LoRA module that maintains independent LoRAs for different users and a Long-Short Modality Retriever that retrieves different history lengths for different modalities, significantly improving performance with negligible time cost increase. Furthermore, we design a Few2Many Learning Strategy, using a conventional recommendation model as a lens to magnify small training spaces to full spaces. Extensive experiments on real-world datasets demonstrate the efficacy of our RecLoRA compared to strong baseline models.

\end{abstract}

\ccsdesc[500]{Information systems~Data mining}
\ccsdesc[500]{Information systems~Recommender systems}

\keywords{Large Language Models, Recommender Systems, User Modeling}

\maketitle

\section{Introduction}

\begin{figure}[t]
  \centering
  \includegraphics[width=0.48\textwidth]{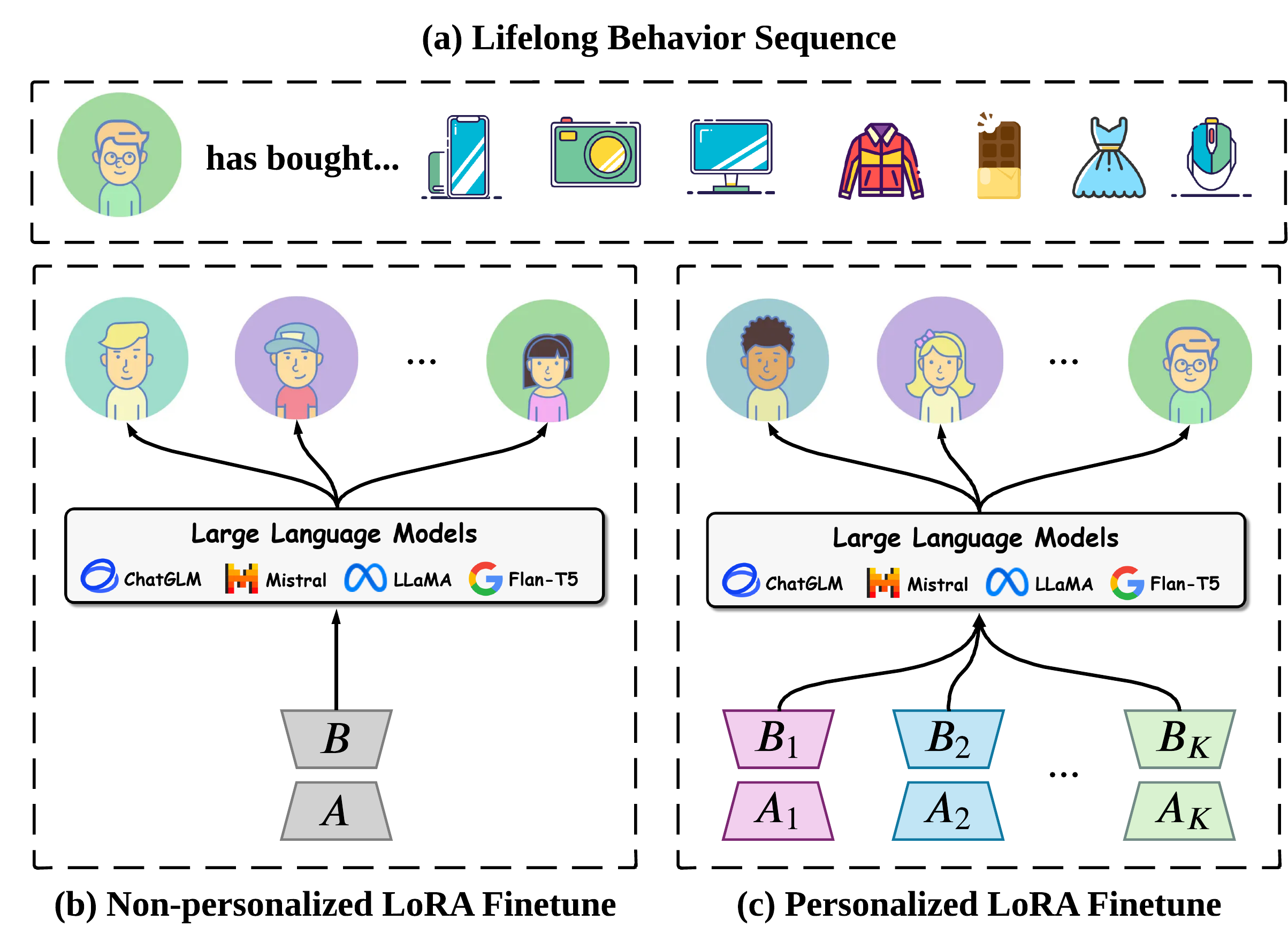}
  \vspace{-10pt}
  \caption{The illustration of lifelong behavior sequence and personalized low-rank adaption (LoRA) of large language models for recommendation.}
  \label{fig:illustration personalized lora}
  \vspace{-10pt}
\end{figure}

Recommender systems play essential roles in various online services to mitigate the information overload problem and meet users' information needs~\cite{guo2017deepfm,lin2023map,liu2024mamba4rec,qin2022rankflow,qin2023learning}. 
Besides, large language models (LLMs) have experienced significant growth in the field of natural language processing (NLP), demonstrating a remarkable ability to comprehend human languages and generate texts that closely resemble human writings for a broad range of tasks~\cite{zhao2023survey,brown2020language,touvron2023llama}. 
Recent studies have started to investigate the potential of LLMs for recommender systems under various recommendation tasks, such as listwise ranking and pointwise scoring~\cite{lin2023can,lin2023rella}.
These studies usually inject personalized knowledge from recommendation domains into LLMs~\cite{lin2023rella,zhang2023collm,zheng2023adapting}.
This personalized knowledge plays a pivotal role in tailoring experiences to individual preferences and enhancing user interactions and engagements~\cite{chen2023large}.
So our crucial points are \textit{what useful personalized knowledge we can provide} and \textit{how to inject it into LLM to help with predictions}.

As the proverb "Actions speak louder than words" suggests, a user's behavior sequence is the most direct expression of their preferences. It typically provides the most detailed personalized information about users in recommendation scenarios.
As illustrated in Figure~\ref{fig:illustration personalized lora}, if a user's behaviors can be collected and
modeled from birth to the present, their interests can be accurately predicted. Several mainstream studies have designed memory mechanisms to model lifelong behavior sequences \cite{ren2019lifelong,pi2019practice}. Other approaches employ retrieval methods to extract relevant information from entire lifelong sequences \cite{qin2020user,pi2020search,chang2023twin}. 
These methods prove the effectiveness of user behavior sequences, which also shows potential in LLM-based recommendation models.

Regarding the second point, there are already existing studies on how to incorporate personalized knowledge into large language models (LLMs). Some works inject personalized information from conventional recommendation models (CRMs) into LLMs by customizing input templates at the input layer \cite{zhang2023collm,li2023e4srec,lin2023clickprompt,bao2023tallrec}. Other studies align representations between semantic and recommendation spaces at the prediction layer \cite{li2023ctrl,wang2023flip}. Despite achieving promising results compared to conventional recommendation models like SASRec \cite{kang2018self} and DIN \cite{zhou2018deep}, existing works still face three significant limitations.

Firstly, personalization is not well established when adapting LLMs to recommender systems. For the sake of memory efficiency, mainstream approaches typically adopt low-rank adaptation (LoRA) techniques~\cite{hu2021lora,dettmers2023qlora} and perform parameter-efficient finetuning (PEFT) to inject personalized knowledge from recommendation domains into LLMs~\cite{lin2023rella,zhang2023collm,zheng2023adapting}. 
In the context of PEFT, maintaining a unique LoRA matrix for each user can enable maximum personalization, effectively capturing and adapting to unique user features and changing interests. However, current methods only implement prompt personalization or representation personalization, while the LoRA module, a core component of fine-tuning LLMs, remains static for all users and target items. 
This lack of parameter personalization results in inferior capabilities to effectively manage user dynamics and preference shifts in modern recommender systems.

Secondly, utilizing lifelong personalized behavior sequences in the prompt or representation layer raises issues of effectiveness and efficiency\cite{lin2023rella}. Since LLMs are sensitive to input length and have a limit on input size, they tend to perform poorly on long behavior sequences, resulting in a ceiling on effectiveness. Additionally, as the length of the input behavior sequence 
increases, the time required for training and inference escalates rapidly, leading to efficiency problems.

Thirdly, existing approaches are not scalable for large-scale industrial datasets due to training inefficiency. Even with LoRA techniques, fine-tuning LLMs on recommendation data consumes immense computational resources and time, making it extremely inefficient for large-scale datasets, typically involving millions or even billions of records. Although some studies \cite{lin2023rella} suggest fine-tuning LLMs on a small fraction (e.g., less than 10\%) of the training data to balance recommendation performance and training efficiency, this approach limits the LLMs' exposure to the full training space, resulting in suboptimal performance. Therefore, enabling LLMs to fully perceive the training space while ensuring training efficiency remains an open research question.

To address the aforementioned issues, we propose a novel Personalized \underline{\textbf{Lo}}w-\underline{\textbf{ra}}nk Adaptation Framework of Large Language Models for \underline{\textbf{Rec}}ommendation (\ie, \textbf{RecLoRA}). 
Specifically, we maintain a set of parallel, independent LoRA weights instead of a single static LoRA module as suggested in previous works \cite{lin2023rella,zhang2023collm}. As shown in Figure~\ref{fig:illustration personalized lora}, this personalized LoRA module functions to incorporate personalized user knowledge into large language models, serving as a strategy for parameter personalization. We assign a sequential CRM and adapt the user representation to personalized LoRA weights. More precisely, we use a soft routing method to aggregate meta-LoRA weights guided by the CRM. Instead of extending user behavior sequences in the input prompt, RecLoRA assigns longer sequences in the sequential CRM, significantly improving effectiveness with negligible cost increase. Moreover, the CRM is pre-trained on the full training space, so dynamically generating LoRA weights based on the CRM can act like a magnifier, enlarging the training signal observed from a small fraction of data to the entire data space without increasing LLM training time.

The main contributions are summarized as follows:
\begin{itemize}[leftmargin=10pt]
    \item To the best of our knowledge, we are the first to consider LoRA parameters personalization in the recommendation task. We design a personalized LoRA module that includes a CRM trained on full training space and a soft routing method to aggregate meta-LoRA weights with CRM instructions.
    \item We propose a novel framework, RecLoRA, which realizes a better trade-off between effectiveness and efficiency. RecLoRA has a Few2Many Magnifier where only few-shot training data is needed to efficiently tune LLM and a Long-short Modality Retriever module, which extends the sequence length in the CRM to improve performance with negligible time cost increase.
    \item Our personalized LoRA module is an easy-to-plugin module, which can be used in any LLM-based recommendation model with LoRA modules.
\end{itemize} 

Extensive experiments on three public real-world datasets have been conducted, where RecLoRA achieves consistent superiority over strong baselines.
\section{Preliminary}
\label{sec:preliminary}

In this section, we will formulate the problem and introduce  Low-Rank Adaptation, which is the basis of RecLoRA. 

\subsection{Problem Formulation}

The core task of recommender systems is to estimate users' preference toward target items given a certain context. 
We denote the historical dataset as $\mathcal{D}=\{(x_i,y_i)\}_{i=1}^N$, where $N$ is the number of data instances. 
The input $x_i$ contains features of user profile $u_i$(\eg, age, location, and historical behaviors), item attributes $v_i$ (\eg, category and brand), context information $c_i$ (\eg, time and season) and user history sequence $\mathcal{H}_{u_i}$(\eg, behaviors like click and ratings).
\begin{equation}
\label{input_x}
\begin{aligned}
    x_i &= (u_i, v_i, c_i, \mathcal{H}_{u_i}),\\
\end{aligned}
\end{equation}
where $\mathcal{H}_{u_i} = \{h_1,h_2,\dots,h_{N_{u_i}}\}$ represents $N_{u_i}$ sequential behaviors of user $u_i$.

The label $y_i\in\{1,0\}$ indicates the interaction signal of the user towards the item (\ie, prefer or not). 
\begin{equation}
	y_{i} = 
		\begin{cases}
		    1,~ u_i~ \text{clicks or likes} ~v_i; \\
			0,~ \text{otherwise.} 
		\end{cases}
\end{equation}

According to different input data transformations, there generally exist two different recommendation paradigms: (1) The input of ID modality $x_i^{ID}$ obtained by one-hot encoding for CRMs (2) The input of textual modality $x_i^{text}$ generated by hard prompt template for LLMs as recommenders.

\subsection{Conventional Recommendation Models}
For ID input $x_i^{ID}$, various conventional recommendation models (CRMs) are designed to capture the collaborative patterns for precise user preference estimation from various aspects (\eg, feature interactions~\cite{wang2021dcn,guo2017deepfm}, user behavior modeling~\cite{pi2020search,qin2020user}). 
We give the general formulation of CRMs as follows:
\begin{equation}
\label{CRM overview}
\begin{aligned}
    h_i^{ID} &= \operatorname{CRM}(x_i^{ID}),\\
    \hat{y}_i^{ID} &= \sigma(\operatorname{MLP}(h_i^{ID}))\in (0,1),
\end{aligned}
\end{equation}
where $\sigma(\cdot)$ is the sigmoid function.

\subsection{Large Language Models as Recommenders}
\label{llm as recommenders}
\begin{figure}[h]
  \centering
  \includegraphics[width=0.46\textwidth]{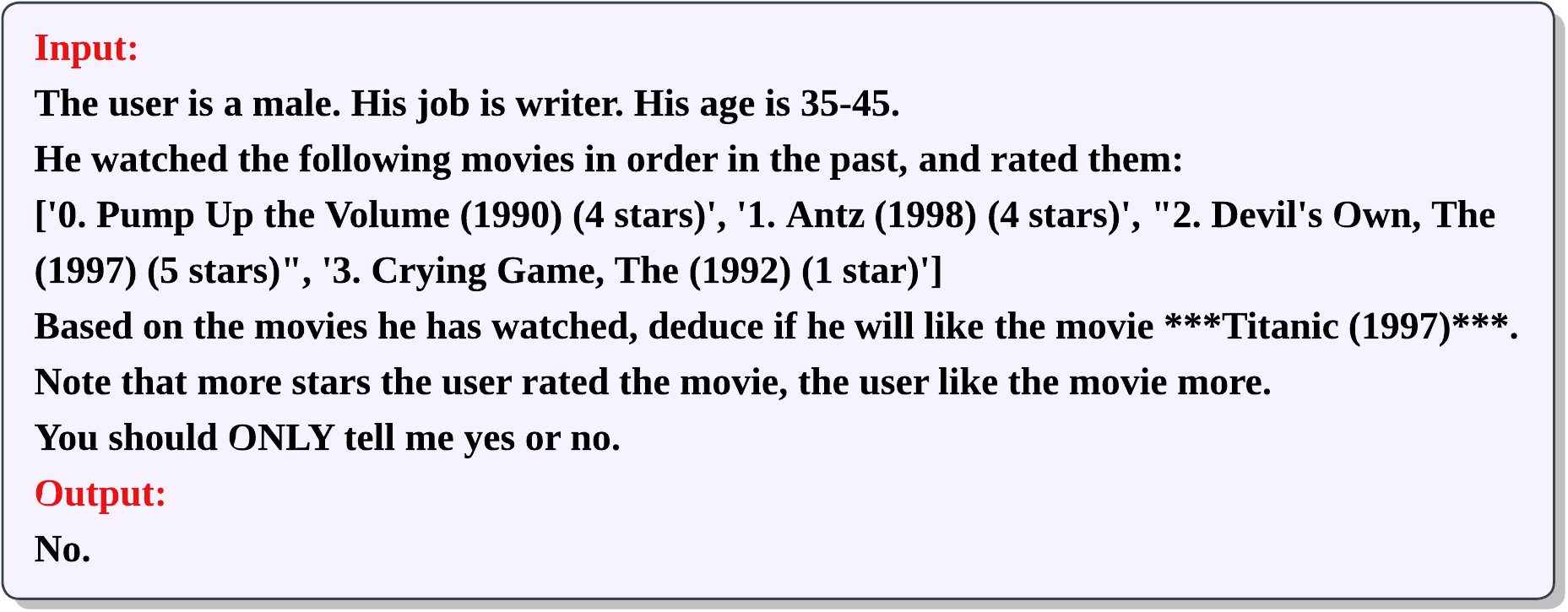}
  \caption{The illustration of textual input-output pair.}
  \vspace{-10pt}
  \label{fig:prompt template example}
\end{figure}

Typically, large language models (LLMs) refer to Transformer-based language models with at least billion-level parameters that are trained on massive text datasets, demonstrating a remarkable capacity in various natural language tasks. 
When adopting directly LLMs as the recommenders, we need to convert the raw data $(x_i,y_i)$ into textual input-output pairs $(x_i^{text},y_i^{text})$ with a hard prompt template. We illustrate one template example in Figure~\ref{fig:prompt template example}.

As shown in Figure~\ref{fig:prompt template example}, the textual input $x_i^{text}$ consists of the user's profile and historical behaviors, followed by a binary question about the personalized preference towards the target item. 
Following previous works~\cite{lin2023rella,bao2023tallrec}, to obtain the floating-point preference estimation $\hat{y}_i^{text}\in[0,1]$ instead of discrete word tokens, we apply bidimensional softmax over the corresponding scores of the binary key answer words (\ie, Yes \& No) from LLMs to accomplish the preference estimation during evaluation:
\begin{equation}
    \hat{y}_i^{text}=\frac{\exp(s_{i,\text{Yes}})}{\exp(s_{i,\text{Yes}})+\exp(s_{i,\text{No}})}\in(0,1).
\end{equation}

\subsection{Low-Rank Adaptation of LLMs}
\label{lora}

Low-rank adaptation (LoRA)~\cite{hu2021lora} serves as a popular parameter-efficient finetuning (PEFT) method to reduce the resource consumption of finetuning LLMs that possess massive parameters.
The basic idea of LoRA is to maintain two trainable lightweight matrices $A$ and $B$ that are attached to a frozen pre-trained weight matrix $W$. 
Hence, the linear transformation $Y=XW$ is reformulated as:
\begin{equation}
    Y^{\prime}=XW+XAB^{T}=X(W+AB^{T}),
\end{equation}
where $X\in\mathbb{R}^{n\times d_{in}}$,\  $W\in\mathbb{R}^{d_{in}\times d_{out}}$,\  $A\in\mathbb{R}^{d_{in}\times r}$,\  $B\in\mathbb{R}^{d_{out}\times r}$, and $r \ll \min\{d_{in},d_{out}\}$. Initially, 
\begin{equation}
    A\sim \mathcal{N}(0,\sigma^2),\;B=0,
\end{equation}
so as to ensure the initial output $Y^{\prime}=XW+XAB^{T}$ equals to $Y=XW$. During fine-tuning, $W$ is fixed while $A$ and $B$ are updated by SGD-based optimization methods like Adam~\cite{kingma2014adam}. 

When applying LoRA to LLM finetuning, we can selectively attach LoRA weights to target matrices of certain types inside the LLM. \eg, the matrices of query, key, and value in the self-attention modules or the linear matrices in feed-forward networks (FFNs).



\section{Methodology}
\label{sec:method}

In this section, we introduce our proposed RecLoRA (Personalized \textbf{\textit{Lo}}w-\textbf{\textit{ra}}nk Adaptation for \textbf{\textit{Rec}}ommendation) framework and its training process in detail.

\begin{figure*}[t]
  \centering
  \vspace{-10pt}
  \includegraphics[width=0.98\textwidth]{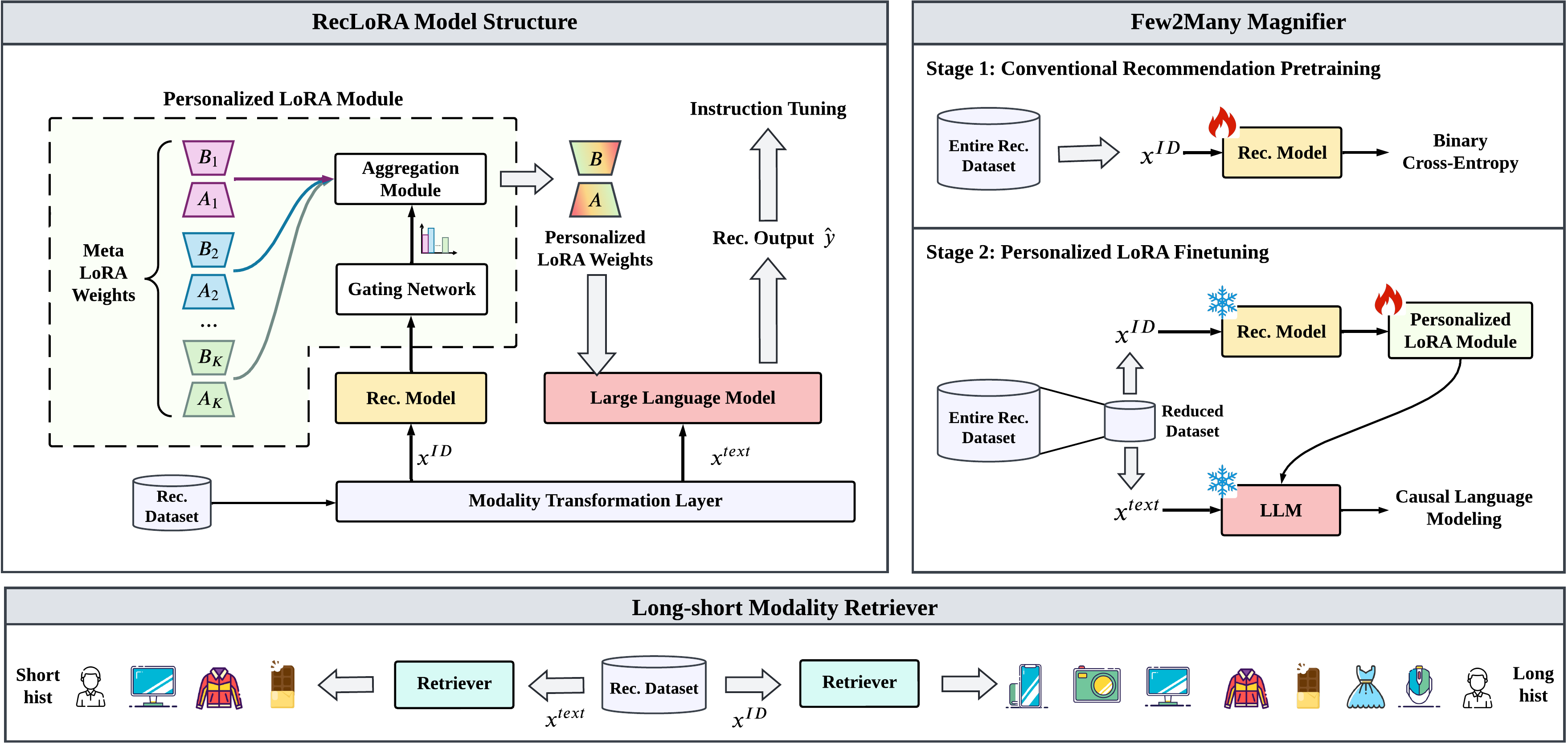}
  \caption{The model structure and learning strategy of our proposed RecLoRA framework. }
  \vspace{-10pt}
  \label{fig:framework}
\end{figure*}

\subsection{Overview of RecLoRA}

We propose the RecLoRA framework to address three major limitations: (1) To construct personalized low-rank adaptation for LLMs in recommendation, we employ a parallel meta-LoRA mechanism incorporating personalized knowledge. (2) To alleviate the effectiveness and efficiency issues associated with behavior sequence extension, we design a Long-Short Modality Retriever. (3) To expand the receptive field of LLMs to the entire training space, we develop a Few2Many Magnifier Strategy. In this section, we provide a comprehensive explanation of our RecLoRA framework, including its network architecture and training process.


Figure~\ref{fig:framework} illustrates the overall framework of the RecLoRA model. As is shown in the figure, RecLoRA takes two input components: samples in ID modality $x^{ID}$ and samples in text modality $x^{text}$. Each modality $x$ has a user profile, a candidate item, context features, and user history behaviors as formulated in Eq.~\eqref{input_x}. So we start by transforming the recommendation dataset into these two modalities and then feed $x^{ID}$ and $x^{text}$ into CRM and LLM models respectively. During transformation, a Long-Short Modality Retriever retrieves long history for $x^{ID}$ and short history for $x^{test}$.
\begin{equation}
\begin{aligned}
    x^{ID} &= LSMR(u^{ID}, v^{ID}),\\
    x^{text} &= LSMR(u^{text}, v^{text}),\\
    N^{ID}_{u_i} &> N^{text}_{u_i},
\end{aligned}
\end{equation}
where $N$ is the sequence length and this module would be described in detail in section~\ref{long-short modality retriever}. 

Once fed with $x^{ID}$ and $x^{text}$, RecLoRA begins its process. Our main predictor for labeling $y$ is the LLM
\begin{equation}
\begin{aligned}
    y &= LLM(x^{text}).\\
\end{aligned}
\end{equation}

Thus, we need to fine-tune the pre-trained LLM with the recommendation dataset. To achieve memory efficiency, we choose to use LoRA for parameter-efficient tuning of the LLM, as introduced in Section~\ref{lora}. However, instead of using traditional LoRA, we design a Personalized LoRA Module to incorporate user-personalized knowledge. This module can be expressed as follows:
\begin{equation}
\label{lora forward}
\begin{aligned}
    h_{j+1} &= Layer_j(h_j)+LoRA(h_j),\\
    h_{j+1} &= Layer_j(h_j)+PLoRA(h_j,x^{ID}),\\
\end{aligned}
\end{equation}
where $Layer_j$ is the $j_{th}$ layer of LLM, and PLoRA is our module, which takes the hidden states from the  $j_{th}$ layer along with $x^{ID}$  as inputs and outputs the personalized representation. The details of PLoRA will be provided in Section~\ref{personalized lora}.

\subsection{Personalized Low-Rank Adaption}
\label{personalized lora}

Intuitively, the most straightforward way for personalized low-rank adaptation is to maintain independent LoRA weights $(A_u,B_u)$ for each user $u$ separately, \ie,
\begin{equation}
    Y^{\prime}_u=XW+XA_u B_u^{T},\;\forall u\in\mathcal{U}.
\end{equation}
However, this approach is non-scalable, as its model complexity grows linearly with the number of users, leading to a tremendous amount of memory consumption. 
Moreover, it overlooks the overlap and commonality among behavior patterns of different users, which can lead to suboptimal performances.

Hence, to balance user-specific individual preference and inter-user pattern commonalities, we propose a Personalized LoRA Module.
Specifically, as shown in Figure~\ref{fig:framework}, we maintain a set of meta-LoRA weights $\{(A_k,B_k)\}_{k=1}^{N_m}$, which are designed to capture the diverse user behavior patterns. 
Each meta-LoRA weight $(A_k,B_k)$ can be regarded as a latent subspace to boost the expressiveness and capacity. 


Then, we can obtain the personalized LoRA weights by combining them with a trainable gating network as
\begin{equation}
\label{weighted sum}
    A_pB_p^T = \sum_{k=1}^{N_m} \alpha_k~A_kB_k^T,
\end{equation}
where $A_p, B_p$ are the final personalized LoRA matrices, and $N_m$ is the number of meta-LoRA weights. 

In this equation, the $k_{th}$ gating weight $\alpha_k$ represents the importance and relevance of the current user and the $k_{th}$ LoRA weight, acting as an indicator for personalization. Thus, to capture the complicated correlations and achieve personalized aggregation of the meta-LoRA weights, we introduce the conventional recommendation model (CRM) to provide personalized representations for the gating network. CRM takes inputs of discrete ID and generates hidden states as personalized representations. 
\begin{equation}
    R_{c} = CRM(x^{ID}),
\end{equation}
where $R_c \in \mathbb{R}^{n\times d_{c}}$ represents the CRM's output. Specifically, we use a sequential behavior model, SIM\cite{SIM}, as user history provides the most ideal personalization, and SIM is a state-of-the-art model for sequential behaviors in recommendations.

After obtaining the representation, we adapt it to the gating weight to guide the meta-LoRA weights with personalized knowledge. We use a feed-forward network (FFN) with a ReLU activation and a LayerNorm as the adapter:
\begin{equation}
\begin{aligned}
    \beta &= W_2 \times \text{ReLU}(\text{LN}(W_1 R_c +b_1)))+b_2,\\
\end{aligned}
\end{equation}
and a softmax function to generate gating weights:
\begin{equation}
\begin{aligned}
    \alpha_j &= \frac{\exp(\frac{\beta_{j}}{tp})}{\sum_{j'=1}^{N_m} \exp(\frac{\beta_{j'}}{tp})} ,
\end{aligned}
\end{equation}
where $W_1 \in \mathbb{R}^{d_c\times d_h}$, $W_2 \in \mathbb{R}^{d_h\times d_r}$, $b_1 \in \mathbb{R}^{d_h}$, $b_2 \in \mathbb{R}^{d_r}$, and $\alpha \in \mathbb{R}^{n\times N_m}$ is the gating weight of Eq.~\eqref{weighted sum}. $tp$ is a temperature hyperparameter in the softmax function, making it smoother or sharper. Most importantly, softmax guarantees 
the distribution of personalized LoRA weights matches each meta-LoRA weight. 

Finally, we obtain the personalized LoRA matrix by Eq.~\eqref{weighted sum} and use it as the traditional LoRA method does in Eq.~\eqref{lora forward}. In this way, this module dynamically generates one unique LoRA module personally for each user, realizing LoRA parameters personalization.

\subsection{Few2Many Magnifier Strategy}
During LLM fine-tuning, the ideal scenario would be exposing the entire recommendation dataset to the LLM, as studies have shown that the more data an LLM is exposed to, the better its performance. However, constraints on time and efficiency make it impractical to expose large datasets on the input side. Therefore, a more efficient approach is to inject large data knowledge without significantly increasing time costs. To achieve this, we propose the Few2Many Magnifier Strategy.

In the first stage, we train a conventional recommendation model using the full training dataset to obtain a well-trained model capable of providing sample-level personalized vector representations with high generalization based on the inputs $x^{ID}$ in ID modality
. This is achieved through a binary cross-entropy loss
\begin{equation}
    \mathcal{L}^{ID} = \sum_{x_i^{ID} \in \mathcal{D}} [-y_{i}^{ID} \cdot \log(\sigma(\hat{y}_{i}^{ID})) - (1 - y_{i}^{ID}) \cdot \log(1 - \sigma(\hat{y}_{i}^{ID}))],
\end{equation}
where $\sigma$ is the sigmoid function, $y_i^{ID}$ is the true label and $\hat{y}_i^{ID}$ is 
the prediction label calculated in Eq.~\eqref{CRM overview}
Now, the CRM has fit the full training space and has learned most of the knowledge contained in the whole dataset.

In the second stage, we downsample the large-scale training set to obtain a smaller training set for the efficient parameter tuning of RecLoRA, which includes large language models. During this process, the fully trained traditional recommendation model from the first stage provides personalized information, ensuring full spatial perspectives. This information is combined to form a personalized LoRA matrix with sufficient generalization, as detailedly described in Section~\ref{personalized lora}.
Thus, even though the large language model sees only a small number of training samples during fine-tuning, it extends its receptive field to the full training space through the conventional recommendation models and personalized LoRA matrices. This greatly enhances the sample efficiency and recommendation performance of the large language model. It should be noted that during this process, only the personalized LoRA module is updated, while the traditional recommendation models and large language models remain fixed. The training loss used is the cross-entropy loss in traditional causal language modeling.

During the inference stage, the large language model does not directly give a pointwise score $\hat{y}^{text} \in \{0,1\}$. Therefore, we intercept the vocabulary scores, and then conduct a bidimensional softmax over the scores of binary key answer words. Specifically, the scores for “Yes”
 and “No” are $s_y$ and $s_n$ respectively. Then, the pointwise scoring of LLMs can be written as: 
\begin{equation}
\begin{aligned}
    \hat{y}^{text} = \frac{\text{exp}(s_y)}{\text{exp}(s_y)+\text{exp}(s_n)}.
\end{aligned}
\end{equation}
This prediction will be used to calculate evaluation metrics.

\subsection{Long-Short Modality Retriever}
\label{long-short modality retriever}
Intuitively, ideal personalized data involves a lifelong behavior sequence. However, the time cost escalates significantly with extended behavior sequences. To address efficiency issues while improving the performance, we propose a Long-Short Modality Retriever.

It has been widely demonstrated that retrieval is effective in extracting important information from long behavior sequences for the current sample prediction \cite{qin2020user,SIM}. However, the history length remains a trade-off between effectiveness and efficiency. Therefore, we retrieve long histories for the input $x^{ID}$  in the CRM and short histories for the input $x^{text}$ in the LLM. 
The CRM, trained on longer histories, can provide more sequential information and personalized knowledge for the LLM through LoRA parameters. This approach hardly increases the time cost, achieving a fantastic balance between effectiveness and efficiency, as will be shown in Section~\ref{sec:time efficiency}.

The retrieval method can be diverse and flexible. For example, we retrieve behaviors using semantic behavior encoding. Specifically, we input each behavior into the LLM and obtain the hidden states from the last layer of the LLM as its representation. We then apply principal component analysis (PCA) for dimension reduction and denoising. Finally, we calculate cosine similarities to identify the top-k most relevant behaviors to the current item.

\section{Experiments}
In this section, we conduct the experimental settings and results. Five research questions lead the following discussions, and our implementation code of RecLoRA is publicly available.\footnote{https://anonymous.4open.science/r/RecLoRA-955D/}

\begin{itemize}
    \item[\textbf{RQ1}] Does RecLoRA outperform existing baselines?
    \item[\textbf{RQ2}] What are the influences of different components in RecLoRA?
    \item[\textbf{RQ3}] What is the influence of meta-LoRA number $N_m$ in Eq.~\eqref{weighted sum}?
    \item[\textbf{RQ4}] How do the performance and time cost increase when extending a longer behavior sequence? 
    \item[\textbf{RQ5}] How does RecLoRA improve the sample efficiency for LLM? 
\end{itemize}
\begin{table}
    \vspace{-5pt}
    \caption{The dataset statistics.}
    \vspace{-10pt}
    \centering
    \resizebox{0.45\textwidth}{!}{
    \renewcommand\arraystretch{1.1}
    \begin{tabular}{c|cccccc}
    \toprule
     Dataset   & \#Users & \#Items & \#Samples & \#Fields & \#Features \\ 
     \midrule
     MovieLens-25M & 162,541 & 59,047 & 25,000,095 & 6 & 280,576 \\ 
     MovieLens-1M &  6,040 & 3,706 &  970,009 & 10 &16,944\\
     GoodReads & 449,114 & 1,432,348 & 20,122,040 & 15 & 5,787,895\\
     \bottomrule
    \end{tabular}
    }
    \vspace{-10pt}
    \label{tab:datasets}
\end{table}

\begin{table*}
\vspace{-10pt}
\caption{The performance of different models in \emph{ID-based} and \emph{LM-based} settings. 
The best result is given in bold, and the second-best value is underlined. 
The symbol $\ast$ indicates a statistically significant improvement of RecLoRA over the best baseline with $p$-value < 0.01.
}
\vspace{-5pt}
\label{tab:All performance}
\resizebox{0.9\textwidth}{!}{
\renewcommand\arraystretch{1.1}
\begin{tabular}{c|c|ccc|ccc|ccc}
\toprule
\hline

\multicolumn{2}{c|}{\multirow{2}{*}{Model}} & \multicolumn{3}{c|}{MovieLens-25M} &\multicolumn{3}{c|}{MovieLens-1M} &\multicolumn{3}{c}{GoodReads}\\ 
\multicolumn{2}{c|}{} & AUC  & Log Loss & Rel.Impr & AUC  & Log Loss & Rel.Impr & AUC  & Log Loss & Rel.Impr\\ 
   \hline

\multicolumn{1}{c|}{\multirow{8}{*}{ID-based}} & DeepFM & 0.8181 & 0.4863 & 3.52\% & 0.7978 & 0.5405 & 2.04\% & 0.7873 & 0.5027 & 1.47\% \\ 
\multicolumn{1}{c|}{\multirow{4}{*}{}} & AutoInt  & 0.8138 & 0.4942 & 3.77\%& 0.7976 & 0.5398 & 2.06\% & 0.7866 & 0.5024 & 1.56\% \\ 
\multicolumn{1}{c|}{\multirow{4}{*}{}} & DCNv2  & 0.8184 & 0.4905 & 3.50\%& 0.7977 & 0.5403 & 2.05\% & 0.7867 & 0.5022 & 1.55\%\\ 
\multicolumn{1}{c|}{\multirow{4}{*}{}} & GRU4Rec  & 0.8169 & 0.4917 & 3.55\%& 0.7959 & 0.5423 & 2.29\% & 0.7869 & 0.5020 & 1.52\% \\ 
\multicolumn{1}{c|}{\multirow{4}{*}{}} & Caser  & 0.8167 & 0.4917 & 3.39\% & 0.7952 & 0.5425 & 2.37\% & 0.7872 & 0.5023 & 1.49\%\\ 
\multicolumn{1}{c|}{\multirow{4}{*}{}} & SASRec  & 0.8163 & 0.4892 & 3.54\% & 0.7984 & 0.5388 & 1.97\% & 0.7864 & 0.5022 & 1.59\%\\ 
\multicolumn{1}{c|}{\multirow{4}{*}{}} & DIN & 0.8258 & 0.4775 & 3.50\% & 0.7989 & 0.5389 & 1.90\% & 0.7880 & 0.4999 & 1.38\%\\ 
\multicolumn{1}{c|}{\multirow{4}{*}{}} & SIM  & 0.8379 & 0.4664  & 1.59\% & 0.7992 & 0.5387 & 1.86\% & \underline{0.7896} & \underline{0.4993} & 1.18\%\\ 
\hline
\multicolumn{1}{c|}{\multirow{4}{*}{LM-based}}  & CTR-BERT  & 0.8079 & 0.5044 & 4.93\% & 0.7931 & 0.5457 & 2.64\% & 0.7176 & 0.5576 & 11.32\%\\ 
\multicolumn{1}{c|}{\multirow{4}{*}{}} & TallRec  & 0.8324 & 0.4733 & 1.65\%  & 0.7789 & 0.5580 & 4.51\% & 0.7759 & 0.5148 & 2.96\%\\ 
\multicolumn{1}{c|}{\multirow{4}{*}{}}& ReLLa  & \underline{0.8409} & \underline{0.4662} & 1.22\% & \underline{0.8005} & \underline{0.5364} & 1.70\% & 0.7833 & 0.5065 & 1.99\% \\ \multicolumn{1}{c|}{\multirow{2}{*}{}} & RecLoRA  & \textbf{0.8462$^*$} & \textbf{0.4552$^*$} &- & \textbf{0.8141$^*$} & \textbf{0.5248$^*$} & - & \textbf{0.7989$^*$} & \textbf{0.4916$^*$}  & -\\ 
  
   \hline  
   \bottomrule          
\end{tabular}
\vspace{-5pt}
}
\end{table*}

\begin{figure*}[t]
  \centering
  \vspace{-5pt}
  \includegraphics[width=1.0\textwidth]{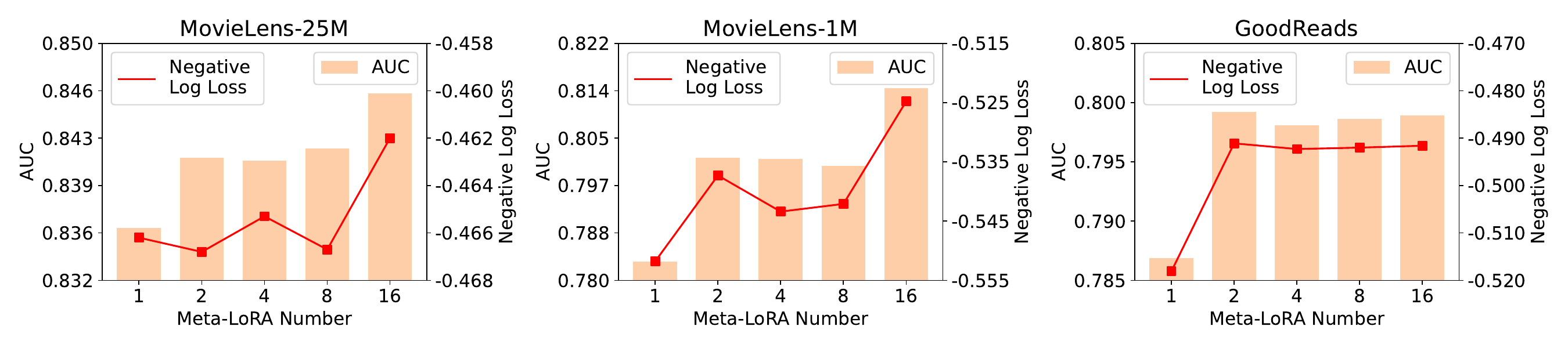}
  \vspace{-20pt}
  \caption{The AUC performance of RecLoRA w.r.t. different meta-LoRA numbers $N_m$. RecLoRA manages to have already performed very well starting from $N_m=2$.}
  \label{fig:meta lora number}
\end{figure*}

\subsection{Experiment Setup}
\subsubsection{Datasets}
We conduct experiments on three real-world datasets (\ie, 
GoodReads\footnote{\url{https://mengtingwan.github.io/data/goodreads.html}}, MovieLens-1M\footnote{\url{https://grouplens.org/datasets/movielens/1m/}} and MovieLens-25M\footnote{\url{https://grouplens.org/datasets/movielens/25m/}}). 
We show the dataset statistics in Table~\ref{tab:datasets} and give detailed data preprocessing information in Appendix~\ref{app:dataset} due to page limitations.

\subsubsection{Evaluation Metrics}
To assess the effectiveness of our methods, we use AUC (Area Under the ROC Curve), Log Loss (binary cross-entropy loss), and Rel.Impr (Relative Improvement) as evaluation metrics. In the context of CTR prediction, even minor variations, such as a 0.001 increase in AUC or a 0.001 decrease in Log Loss, are considered significant improvements \cite{xDeepFM, DCNv2}.

\subsubsection{Baseline Models}

We have divided our baselines into two main categories:(1) \emph{traditional ID-based}, containing DeepFM \cite{DeepFM}, AutoInt \cite{AutoInt}, and DCNv2 \cite{DCNv2} as exemplars of feature interaction models and  GRU4Rec \cite{GRU4Rec}, Caser \cite{Caser}, SASRec \cite{SASRec}, DIN \cite{zhou2018deep}, and SIM \cite{SIM} as key representatives of user behavior models (2) \emph{LM-based}, containing CTR-BERT\cite{CTRBERT}, TALLRec \cite{bao2023tallrec}, and ReLLa \cite{lin2023rella}. Detailed descriptions of these baselines are provided in Appendix~\ref{app:baseline}.

\subsubsection{Implementation Details} We have provided our details, such as the base model, hyper-parameters, and prompts, in Appendix~\ref{app:details}.

\subsection{Overall Performance (RQ1)}
We evaluate the performance of RecLoRA in comparison to existing baseline models, with the results reported in Table~\ref{tab:All performance}. It is important to note that most of the other recommendation baseline models are trained in full-shot settings with the entire training set, while only TallRec, ReLLa and RecLoRA are trained on few-shot training sets with 70,000 (<10\%) training samples in all three datasets. We set the length of the text and ID behavior sequences to 10 and 60, respectively, in all three datasets. \emph{ReI.Impr} denotes the relative AUC improvement rate of RecLoRA against each baseline. Our observations from Table~\ref{tab:All performance} are as follows.
 \begin{itemize}[leftmargin=10pt]
     \item SIM achieves the best performance among all the ID-based baseline models. By applying user behavior retrieval, SIM effectively reduces the noise in user sequences, which is fundamentally beneficial for CTR predictions. Additionally, the traditional LM-based CTR model CTR-BERT performs worse than most ID-based traditional CTR models, consistent with the results reported in [40, 65]. It only incorporates the small language model BERT [2] for pure text-based recommendations, resulting in an inferior performance.
     \item ReLLa generally achieves a performance comparable to the best ID-based baseline model, SIM. ReLLa incorporates retrieval-enhanced instruction tuning (ReiT) and utilizes only 10\% of the training data. This observation demonstrates the effectiveness of both tuning LLM and retrieval in the input sequences.
     \item RecLoRA outperforms the state-of-the-art baselines significantly with a $p$-value < 0.01 against the best baseline, validating the effectiveness of our proposed Personalized LoRA module. The better performance with only 10\% of the training data for tuning LLM demonstrates that our RecLoRA model effectively captures user interests and achieves superior user personalization.
 \end{itemize}

\subsection{Ablation Study (RQ2)}
In this section, we conduct ablation experiments to analyze the effectiveness of RecLoRA's main components: the Personalized LoRA and the Long-Short Modality Retriever.


\begin{itemize}[leftmargin=10pt]
     \item \textbf{RecLoRA (Ours)}: This is the complete version of our proposed method.
     \item \textbf{RecLoRA (w/o meta-LoRA)}: We remove the Personalized LoRA module and only maintain a LoRA module with a rank of 8.
     \item \textbf{RecLoRA (w/o meta-LoRA same param)}: Since the personalized LoRA module has $N_m=16$ times more training parameters, we maintain the parameter count by replacing the meta-LoRA weights with a single LoRA weight, increasing the LoRA rank from 8 to 128.
     \item \textbf{RecLoRA (w/o long retriever)}: To evaluate the effect of retrieval history in $x^{ID}$, we remove the long retriever module and only use the recent history.
     \item \textbf{RecLoRA (w/o short retriever)}: To evaluate the effect of retrieval history in $x^{text}$, we remove the short retriever module and only use the recent history.
     \item \textbf{RecLoRA (w/o long-short retriever)}: To evaluate the overall effect of retrieval history, we remove both the long and short retrievers, using only the recent history.
\end{itemize}

The results are shown in Table~\ref{tab:w/o meta lora}. We observe that the performance of RecLoRA significantly decreases when the personalized LoRA module is removed. This finding confirms that the lack of user personalization knowledge in the LLM tuning process leads to suboptimal performance. Our proposed personalized LoRA module effectively addresses this issue and improves performance. Additionally, we observe from the third line that the improvement is not merely due to an increase in the parameter count by $N_m$ times. 

Meanwhile, we observe that removing either the long or short retriever generally results in a performance drop. This highlights the importance of considering lifelong user sequences and demonstrates that inputs from different modalities both benefit from retrieving long sequences.

\begin{table*}
    \vspace{-5pt}
    \caption{The performance of different variants of RecLoRA. We remove different components of RecLoRA to evaluate the contribution of each part to the model. The best result is given in bold, and the second-best value is underlined..}
    \vspace{-5pt}
    \centering
    \resizebox{1.0\textwidth}{!}{
    \renewcommand\arraystretch{1.1}
    \begin{tabular}{c|ccc|ccc|ccc}
    \toprule
    \multirow{2}{*}{Model} & \multicolumn{3}{c|}{ML-25M} & \multicolumn{3}{c|}{ML-1M} & \multicolumn{3}{c}{GoodReads} \\ 
    \multirow{2}{*}{} & AUC  & Log Loss & Rel.Decr & AUC  & Log Loss & Rel.Decr & AUC  & Log Loss & Rel.Decr \\ 
     \midrule

     RecLoRA & 0.8462 & \textbf{0.4552} & - & \textbf{0.8141} & \textbf{0.5248} & - &\textbf{0.7989} & \textbf{0.4916} & -\\
     RecLoRA (w/o meta-LoRA) & 0.8360 & 0.4662 & -1.21\% & 0.7869 & 0.5517 & -3.34\% & 0.7740 & 0.5181 & -3.12\% \\
     RecLoRA (w/o meta-LoRA same param) & 0.8370 & 0.4759 & -1.09\% & 0.7820 & 0.5548 & -3.94\% & 0.7735 & 0.5125& -3.17\% \\
     RecLoRA (w/o long retriever) & \textbf{0.8501} & 0.4761 & +0.46\% & \underline{0.8026} & \underline{0.5321} & -1.41\% &0.6744 & 0.6231 & -15.6\%\\
     RecLoRA (w/o short retriever) & \underline{0.8473} & \underline{0.4606} & +0.13\% & 0.8001 & 0.5359 & -1.71\%& \underline{0.7971} & \underline{0.4924} & -0.22\%\\
     RecLoRA (w/o long-short retriever) & 0.8453 &  0.4787 & -0.11\% & 0.7978 & 0.5373 & -2.00\% &0.6882 & 0.6073 & -13.9\%\\
     \bottomrule
    \end{tabular}
    }
    \vspace{-5pt}
    \label{tab:w/o meta lora}
\end{table*}

\begin{table}
    \caption{The performance and time cost of ReLLa and RecLoRA in GoodReads w.r.t. different history length in text and ID inputs.}
    \vspace{-5pt}
    \centering
    \resizebox{0.48\textwidth}{!}{
    \renewcommand\arraystretch{1.1}
    \begin{tabular}{cc|cc|cc|cc}
    \toprule
    \multicolumn{2}{c|}{Hist Length} & \multicolumn{2}{c|}{AUC} &\multicolumn{2}{c|}{Training}&\multicolumn{2}{c}{Inference} \\ 
      Text & ID & Value & Improve & Time & Increase & Time & Increase\\ 
     \midrule

     10 & -  &0.7776 & - & 1.4s/sample & - & 0.6s/sample  & -\\
     30 & -  &0.7828 & +0.6\% & 3.34s/sample&+138.78\% & 1.06s/sample   &+76.67\%\\
     60 & - & \underline{0.7833} & +0.73\%& 6.41s/sample &+358.16\%& 1.83s/sample   & +205.55\% \\
     10 & 30 & 0.7822 & +0.59\% & 1.42s/sample &+2.04\% & 0.63s/sample & +5.55\%\\
     10 & 60 & \textbf{0.7989} & +2.73\% & 1.45s/sample & +4.08\% & 0.65s/sample   &+8.33\%\\
     \bottomrule
    \end{tabular}
    }
    \vspace{-10pt}
    \label{tab:time efficiency}
\end{table}
\begin{figure*}[t]
  \centering
  \includegraphics[width=1.0\textwidth]{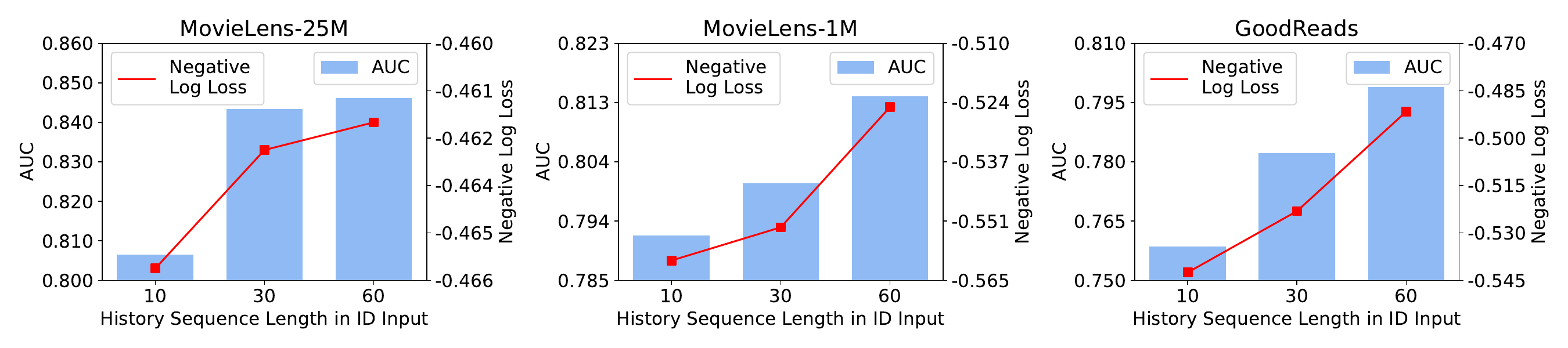}
  \vspace{-20pt}
  \caption{The performance of RecLoRA w.r.t. different history length in ID inputs.}
  \label{fig:hist length}
  \vspace{-10pt}
\end{figure*}


\subsection{Hyperparameter Study (RQ3)}
In this section, we conduct hyperparameter experiments to analyze the influence of the meta-LoRA number $N_m$ in the personalized LoRA module. The results are shown in Figure~\ref{fig:meta lora number}.

As shown in Figure~\ref{fig:meta lora number}, the evaluation metric AUC generally increases with the hyperparameter $N_m$, reaching a maximum value at $N_m=16$. This experiment demonstrates that increasing the parameter count improves overall performance, as the hyperparameter $N_m$ determines the number of meta-LoRA weights and the associated trainable parameters.

Furthermore, when $N_m=1$, the personalized LoRA module collapses into a single LoRA module without CRM instruction, resulting in the worst performance. Notably, even starting from $N_m=2$, RecLoRA performs very well, even comparable to $N_m=16$ in GoodReads. This illustrates the effectiveness of personalized LoRA without significantly increasing the parameter count.

\subsection{Time Efficiency (RQ4)}
\label{sec:time efficiency}
In this section, we conduct time efficiency experiments to analyze the balance of performance and time cost when increasing sequence lengths in the RecLoRA model. The results of GoodReads are shown in Figure~\ref{fig:hist length} and Table~\ref{tab:time efficiency}. The results of the other two datasets are similar and omitted in Table~\ref{tab:time efficiency} due to page limits.

Figure~\ref{fig:hist length} displays the performance of RecLoRA as the ID sequence length increases. We observe that the performance improves significantly with longer ID sequences. Since CRM provides instructions for meta-LoRA weights aggregation in the personalized LoRA module, a better performance of CRM theoretically enhances RecLoRA's performance, which is consistent with our experimental results.

Table~\ref{tab:time efficiency} shows the performance and time cost of RecLoRA with different historical lengths in text and ID inputs. It reveals that while increasing text sequence lengths improves performance by +0.73\%, it also leads to a substantial and unacceptable increase in training and inference time costs of +358.16\% and +205.55\%, respectively. In contrast, not only the performance is improved significantly by +2.73\% when extending ID history lengths, but also the corresponding training and inference time only increase by +4.04\% and +8.33\%, respectively, which are negligible compared to the time increases associated with longer text sequences. These results highlight RecLoRA's advantage in balancing effectiveness and efficiency.

\begin{figure*}[t]
  \centering
  \hspace{-20pt}
  \vspace{-10pt}
  \includegraphics[width=1.0\textwidth]{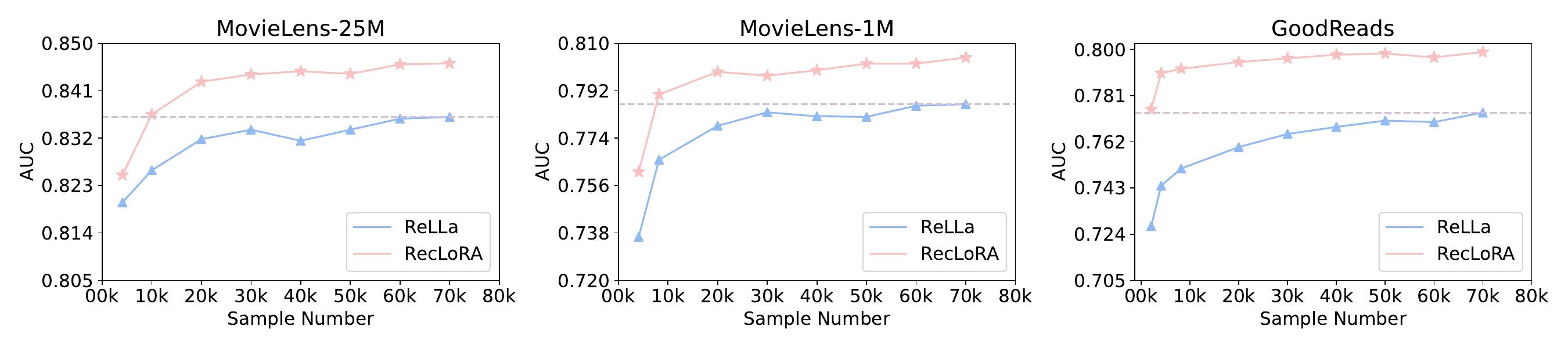}
  \caption{The AUC performance of ReLLa and RecLoRA models w.r.t different sample numbers used to tune LLM. RecLoRA obviously has a higher sample efficiency compared to ReLLa.}
  \vspace{-10pt}
  \label{fig:sample efficiency}
\end{figure*}

\subsection{Sample Efficiency (RQ5)}

In this few-shot setting, we want to explore how much data is necessary for LLMs to adapt to CTR distributions. So we investigate the sample efficiency by varying the number of samples used in training LLM, following previous works\cite{lin2023rella}. In Figure~\ref{fig:sample efficiency}, we report the AUC performance of ReLLa (a few-shot baseline) and RecLoRA with different sample sizes (ranging from 2,048 to 70,000). The text and ID behavior sequence lengths are set to 10 and 60, respectively.

As depicted in Figure~\ref{fig:sample efficiency}, both ReLLa and RecLoRA show performance enhancements as sample numbers gradually increase. However, with the same sample number, RecLoRA consistently outperforms ReLLa by a significant margin. Moreover, even with a small sample size (10,000), RecLoRA performs better than ReLLa with a relatively large sample size (70,000). This is due to RecLoRA's Few2Many Strategy, which utilizes a fully pre-trained CRM for instruction, helping LLMs adapt to CTR distributions better with fewer samples. 

With a limited number of training samples, RecLoRA demonstrates remarkable sample efficiency and considerable few-shot inference ability, benefiting from the mutual assistance of the open-world knowledge of LLMs and the collaborative signals of CRMs.

\section{Related Works}

\subsection{Large Language Model for Recommendation}

Previous work~\cite{lin2023can} has suggested that the employment of language models to recommender systems can generally be categorized based on the roles they play in the recommendation pipeline. \ie, feature engineering~\cite{liu2023first,borisov2022language,li2023taggpt,mysore2023large,carranza2023privacy,christakopoulou2023large}, feature encoder~\cite{muhamed2021ctr,hou2022towards,yu2021tiny,wang2022transrec,hou2023learning,zhang2022twhin,fu2023exploring,yuan2023go,qiu2021u,li2023exploring,he2022ptm4tag,muhamed2021ctr}, scoring/ranking function~\cite{liu2022ptab,kang2023llms,zhang2021language,li2023pbnr,bao2023tallrec,li2023text,zhang2023prompt,mao2023unitrec,hua2023up5,geng2023vip5,hua2023index,zhang2023chatgpt,hou2023large,chen2023palr,petrov2023generative,wang2023zero}.

In feature engineering, large language models (LLMs) process raw data (\eg, user profiles and item descriptions) as input, and output open-world knowledge or potential new attributes for data augmentation with carefully designed prompts or templates.
For example, KAR~\cite{xi2023towards} utilizes the potential knowledge of user preferences and item attributes by requesting LLMs with factorization prompting techniques. 
GENRE~\cite{liu2023first} employs LLMs to generate news summarization, synthetic pieces, and user profiles. 
The obtained knowledge serves as augmented features and improves performances of recommenders in a model-agnostic manner. 

In feature encoders, LLMs are used as auxiliary textual encoders to both enrich the user/item representations with semantic information and enable cross-domain recommendation with the open-world natural language interface. 
For instance, U-BERT~\cite{qiu2021u} enhances user representation by encoding review texts into dense vectors via BERT. UniSRec~\cite{hou2022towards} and VQ-Rec~\cite{hou2023learning} apply a fixed BERT as the encoder for item descriptive texts, to achieve unified cross-domain sequential recommendation.

In scoring/ranking function, instead of doing assistant tasks for recommendation (\eg, feature engineering or feature encoder), LLMs are adopted to do the scoring or ranking task, which is the core component of recommendation.
In this case, LLMs try to accomplish either the item scoring task~\cite{liu2022ptab,kang2023llms,zhang2021language,li2023pbnr,bao2023tallrec,li2023text,zhang2023prompt,mao2023unitrec,lin2023rella}, or item generation task~\cite{hua2023up5,geng2023vip5,hua2023index,zhang2023chatgpt,hou2023large,chen2023palr,petrov2023generative,wang2023zero}. 
Also, various works~\cite{geng2022recommendation,cui2022m6,zhang2023recommendation,liu2023chatgpt,sun2023chatgpt,dai2023uncovering} attempt to explore the multi-task capacity of LLMs, and instruct LLMs with various ways to solve the multiple tasks (\eg, both scoring and generation) through a unified language interface.

In this paper, we focus on utilizing LLMs as scoring functions, using LoRA techniques to parameter-efficiently tune LLM to fit the CTR distribution. We design a personalized LoRA module, which aggregates LoRA weights guided by conventional recommendation models. To the best of our knowledge, we are the first to consider LoRA parameters personalization by introducing CRM instructions in the large language model for recommendation setting. 

\subsection{Long Sequence User Modeling}
During recommendation task, Kim~\cite{kim2003learning} argues that long-term sequence means general interest, which is back to one’s mind and important for personalization. Existing approaches for addressing long-term user sequence mainly focus on memory network and retrieval methods. Hierarchical Periodic Memory Network (HPMN)~\cite{ren2019lifelong} proposes a hierarchical and periodical updating mechanism for capturing multi-scale sequential user interests. The Memory Augmented Neural Network (MIMN)~\cite{pi2019practice} stores behaviors in a memory matrix at the user interest center and updates the memory for new users.

Sequential Interest Modeling (SIM)~\cite{pi2020search} and User Behavior Retrieval for CTR (UBR4CTR)~\cite{qin2020user} have introduced retrieval-enhanced history and two-stage frameworks to catch user patterns in the past, which are related to current targets. UBR4CTR employs BM25 as the relevance metric in its initial stage. In contrast, the original SIM has two variations with different designs. SIM Hard selects relevant items based on the same category as the target item, while SIM Soft utilizes the inner product of pre-trained item embeddings to determine relevance. Efficient Temporal Attention (ETA)~\cite{chen2021end} employs locality-sensitive hashing (LSH) to encode item embeddings and retrieve relevant items from long sequences via Hamming distance. Sequential Dynamic Interest Modeling (SDIM)~\cite{cao2022sampling}  samples behavior items that share the same hash signature as the target item through multiple rounds of hash collisions, and then linearly aggregates these sampled behavior items to capture user interests.

Inspired by these works above, we propose a novel Long-Short Modality Retriever to address long sequence problems in LLM-based recommendation, showing both its effectiveness and efficiency.


\section{Conclusion}
In this paper, we explore the application of large language models in recommendation systems.
We begin by addressing the current challenges related to personalization and long sequences. 
To tackle these issues, we propose RecLoRA, a model designed for lifelong personalized low-rank adaptation. RecLoRA includes a Personalized LoRA module, which maintains distinct LoRAs for individual users.
Additionally, a Long-Short Modality Retriever extracts varying history lengths for different modalities, enhancing performance with negligible time increase. 
We also employ a conventional recommendation model to extend the scope of LLMs to the full training space. 
RecLoRA shows promising performance in offline experiments on three public datasets. 
Further, ablation, hyperparameter, and efficiency studies demonstrate its strong balance between effectiveness and efficiency. 
For future work, we plan to explore more advanced structures for personalization in LLMs and address the fairness issue, which is crucial for achieving complete personalization.


\bibliographystyle{ACM-Reference-Format}
\bibliography{acmart}

\appendix

\section{Data Preprocessing}
\label{app:dataset}
Our experiments are conducted on three real-world public datasets (\ie, MovieLens-25M, MovieLens-1M, GoodReads), and the statistics of the processed datasets are show in Table~\ref{tab:datasets}. All these datasets are split into training, validing and testing sets with ratio of 8:1:1 according to the global timestamp~\cite{qin2021retrieval}. 

\begin{itemize}[leftmargin=10pt]
    \item \textbf{MovieLens-25M} has a scoring range from 0 to 5, with increments of 0.5. We label samples with ratings above 3.0 as positive, and the rest as negative.
    \item \textbf{MovieLens-1M} contains user-movie integer ratings ranging from 0 to 5. Samples with ratings of 4 and 5 are labeled as positive and the rest as negative.
    \item  \textbf{GoodReads} is a book recommendation dataset from GoodReads  website with ratings ranging from 1 to 5. We transform the ratings into binary labels with a threshold of 4. 
\end{itemize}

We apply 5-core filtering to ensure each user or item has at least five interaction records.

Under the few-shot setting with a particular number of training data, we uniformly sample $N$ data instances from the training set, which is then fixed during few-shot tuning.

\section{Baseline Implementation}
\label{app:baseline}

In this section, we describe the baseline models. Baseline models can broadly be divided into two main categories: (1) \emph{traditional CTR models}, which primarily use one-hot encoded IDs as inputs, and (2) \emph{LM-based models}, which integrate pre-trained language models to approach CTR prediction as either a text classification or a sequence-to-sequence problem.

Within traditional CTR models, we distinguish between (1) feature interaction models and (2) user behavior models. We have selected DeepFM \cite{DeepFM}, AutoInt \cite{AutoInt}, and DCNv2 \cite{DCNv2} as exemplars of feature interaction models, and GRU4Rec \cite{GRU4Rec}, Caser \cite{Caser}, SASRec \cite{SASRec}, DIN \cite{zhou2018deep}, and SIM \cite{SIM} as key representatives of user behavior models. For the feature interaction models, we implement average pooling across users' historical behaviors and treat the results as additional feature fields.

SIM is a classic sequential CTR model that employs user behavior retrieval techniques to enhance recommendation performance. We include it in our evaluation to ensure a comprehensive and fair comparison, maintaining consistency with the retrieval method used in RecLoRA.

For LM-based CTR models, we have chosen CTR-BERT \cite{CTRBERT}, TALLRec \cite{bao2023tallrec}, and ReLLa \cite{lin2023rella} as baseline models to represent this category, and they are representatives of traditional language models and large language models respectively.

\subsection{Traditional CTR Models}
We choose the embedding size from \{32, 64\} on three datasets. The dropout rate is selected from \{0.0, 0.1, 0.2\}. The activation function is fixed to ReLU. The learning rate is selected from $1\times 10^{-3}, 5 \times 10^{-4}, 1\times 10^{-4}$ and AdamW~\cite{adamw} optimizer is used. The batch size is selected from \{64,128,256, 512,1024\}. More model-specific hyperparameter settings are shown as follows:

\begin{itemize}[leftmargin=10pt]
    \item \textbf{DeepFM}~\cite{DeepFM}. We choose the size of DNN layer from \{128, 256\} and the number of DNN layers from \{3, 6, 9, 12\}.
    \item \textbf{AutoInt}~\cite{AutoInt}. The attention layers is selected from \{3, 6, 9, 12\} and the attention size is selected from \{64, 128, 256\}. The number of attention heads are all set to 1.
    \item \textbf{DCNv2}~\cite{DCNv2}. We choose the size of DNN layers from \{128, 256\} and the number of DNN layers and cross layers are from \{3, 6, 9, 12\}.
    \item \textbf{GRU4Rec}~\cite{GRU4Rec}. The number of GRU layers is selected from \{1, 2, 3\}. The GRU hidden size and DNN hidden size is selected from \{64, 128, 256\}.
    \item \textbf{Caser}~\cite{Caser}. The number of vertical convolution kernels is selected from \{2, 4, 8\}. The number of horizontal convolution kernels is selected from \{4, 8, 16\}. The number of DNN layers is selected from \{1,2,3\}. The DNN hidden size is selected from \{64, 128, 256\}.
    \item \textbf{SASRec}~\cite{SASRec}. The number of attention heads is selected from \{1, 2, 4\}. The number of attention layers is selected from \{1, 2, 3\}. The attention size is selected from \{64, 128, 256\}. The number of DNN layers is selected from \{1,2,3\}. The DNN hidden size is selected from  \{64, 128, 256\}.
    \item \textbf{DIN}~\cite{zhou2018deep}. The number of DIN attention layers and DNN layers are selected from \{1, 2, 3\}. The DNN hidden size is selected from \{64, 128, 256\}.
    \item \textbf{SIM}~\cite{SIM}. The number of attention layers and DNN layers are selected from \{1, 2, 3\}. The DNN hidden size is selected from\{64, 128, 256\}.
\end{itemize}

\subsection{LM-based Models}
The structure of the pre-trained language models is kept unchanged. And AdamW~\cite{adamw} optimizer is used for all the baselines. The detailed training settings are as follows:

\begin{itemize}[leftmargin=10pt]
    \item \textbf{CTR-BERT}~\cite{CTRBERT}. We maintain a two-tower model structure based on the BERT~\cite{devlin2018bert} model to encode the user and item information respectively. The total number of tuning epochs is set to 10. The batch size is set to 1024. The learning rate is set to $5\times 10^{-5}$ with linear decay. The warmup ratio is 0.05.
   \item \textbf{TallRec}~\cite{bao2023tallrec} is a supervised fine-tuning framework with Llama as a backbone pretrained language model for recommendation task. For fair comparison, we generate the same history in text as ID-based baselines. The total number of tuning epochs is selected from \{10,5\}. The batch size is selected from \{64,128,256\}. The learning rate is selected from \{$1\times 10^{-3}, 5 \times 10^{-4}, 1\times 10^{-4}$\}. 
   \item \textbf{ReLLa}~\cite{lin2023rella} uses user behavior retrieval and supervised fine-tuning with Llama as a backbone pretrained language model for recommendation task. For fair comparison, the retrieval methods in SIM, ReLLa and RecLoRA maintain the same. The total number of tuning epochs is selected from \{10,5\}. The batch size is selected from \{64,128,256\}. The learning rate is selected from \{$1\times 10^{-3}, 5 \times 10^{-4}, 1\times 10^{-4}$\}
   \
\end{itemize}

\section{Implementation Detals}
\label{app:details}

We selected Vicuna-7B \cite{vicuna2023}, provided by FastChat\footnote{\url{https://github.com/lm-sys/FastChat}}, as the foundational large language model (LLM) for our experiments. All computations were performed using V100 GPUs. To enhance training resource efficiency, we employed 8-bit quantization. The LoRA configuration was set with a rank of 8, an alpha value of 16, and dropout set to 0.0. LoRA update matrices were specifically applied to the query and value projection matrices within the attention blocks.

For the instruction tuning process, our detailed configuration such as optimizer and hyperparameters are provided in we utilized the AdamW \cite{adamw} optimizer, configuring it with zero weight decay. The training process was carried out with batch sizes chosen from \(\{64, 128, 256\}\) and initial learning rates selected from \(\{1\times 10^{-3}, 5 \times 10^{-4}, 3 \times 10^{-4}, 1 \times 10^{-4}\}\) using a linear scheduler. For dataset-specific training, the maximum number of epochs was set to 10 for the  MovieLens-25M datasets and 5 for MovieLens-1M and GoodReads datasets. Detailed configurations of the baseline models are provided in Appendix~\ref{app:baseline}. Additionally, the hard prompt templates for textual input-output pairs and item descriptions across all datasets are detailed in Figure~\ref{fig:prompt template example}

Furthermore, when formulating the hard prompt template for RecLoRA, we excluded all pure ID fields—namely, \textit{User ID}, \textit{Book ID} fields in GoodReads, \textit{User ID}, \textit{Movie ID}, and \textit{Zipcode} fields in the MovieLens-1M dataset, as well as
\textit{User ID} and \textit{Movie ID} fields in the MovieLens-25M dataset. This exclusion was based on findings that LLMs exhibit limited perceptual capabilities for pure ID texts~\cite{lin2023can}. In contrast, other fields were utilized as user profile or item information in the prompts, as outlined in Section~\ref{llm as recommenders} and Figure~\ref{fig:prompt template example}. Notably, for other CTR baseline models, we retained all feature fields and user behavior sequences as inputs, ensuring no features were discarded.

\end{document}